\documentclass[12pt]{article}
\usepackage{graphicx}
\usepackage{float}
\usepackage{amsfonts}
\begin{document}

\title{\bf Spherical Symmetric Gravitational Collapse in Chern-Simon Modified Gravity}

\author{M. Jamil Amir\thanks{mjamil.dgk@gmail.com}$~~$and Sarfraz Ali\thanks{sarfraz270@yahoo.com}
\\Department of Mathematics, University of Sargodha, Pakistan,\\
Department of Mathematics, University of Education, Lahore,\\Pakistan.\\}

\date{}

\maketitle

\begin{abstract}
 This paper is devoted to investigate the gravitational collapse in the framework of Chern-Simon (CS) modified gravity.
For this purpose, we assume the spherically symmetric metric as an interior region and the Schwarzchild spacetime
is considered as an exterior region of the star. Junction conditions are used to match the interior and exterior
spacetimes. In dynamical formulation of CS modified gravity, we take the scalar field $\Theta$ as a
function of radial parameter $r$ and obtain the solution of the field equations. There arise two cases where in one case the 
apparent horizon forms first and then singularity while in second case the order of the formation is reversed. 
It means the first case results a black hole which supports the cosmic censorship hypothesis (CCH). Obviously, the second 
case yields  a naked singularity. Further, we use Junction conditions have to calculate the gravitational mass. In non-dynamical 
formulation, the canonical choice of scalar field $\Theta$ is taken and it is shown that the obtained results of CS modified gravity 
simply reduce to those of the general relativity (GR). It is worth mentioning here that the results of dynamical case will reduce to those of GR,
available in literature, if the scalar field is taken to be constant.
\end{abstract}


\section{Introduction}

 Gravitational collapse is one of the most crucial and burning issues in general relativity (GR). Self-gravity of
highly massive stars is a reason behind the occurrence of gravitational collapse. Hawking et al. gave the singularity theorem
\cite{[1]} that there exist spacetime singularities in a generic gravitational collapse. According to
cosmic censorship hypothesis (CCH) \cite{[5]} the singularities appearing in gravitational collapse are
always enveloped by the event horizon. The ultimate result of gravitational collapse depends upon the
choice of initial data and equation of state.

Oppenheimer and Snyder \cite{[3]} pioneerly studied the gravitational dust collapse by considering Friedman-like solution as interior and the Schwarzschild metric as exterior regions. Following them, many researchers
\cite{[4]} have worked on collapse by considering different appropriate geometry of interior and exterior regions.
Misner and Sharp \cite{[6]} extended this work for the perfect fluid. Vaidya \cite{[7]}
used the idea of outgoing radiation of the collapsing body.
Markovic and Shapiro \cite{[9]} generalized the study of collapse with a positive cosmological constant.
Lake \cite{[10]} used positive and negative cosmological constant for the calculation of gravitational collapse.
Sharif and Ahmad \cite{[11]}-\cite{[14]} investigated the spherically symmetric gravitational collapse
with a positive cosmological constant for a perfect fluid. Furthermore, they extended this work for
plane symmetric gravitational collapse using junction conditions \cite{[15]}. Soon after, Sharif and Khadija \cite{[16]}
extended this work using spherical symmetry.

The very interesting and perplexing problem of gravitational collapse has been investigated using alternative
theories of gravity \cite{[17]}-\cite{[21]}  since last decade. There are clear evidences of naked singularities if someone
extend the GR \cite{[17]}. The conditions for the formation of a naked singularity
in the collapse of null dust in higher dimensional f(R) gravity have been obtained in \cite{[22]}.
Openheimer-Snyder collapse in Brans-Dicke theory has been studied in \cite{[23]}. Rudra and Debnath
\cite{[24]} studied gravitational collapse in Vaidya spacetime for Galileon theory of gravity. Sharif and Abbas
\cite{[25]} investigated the dynamics of shearfree dissipative gravitational collapse in f(G) gravity. Spherically
symmetric perfect fluid gravitational collapse has been analyzed in metric f(R) gravity in \cite{[26]}.
In \cite{[27]}, a general f(R) model with uniformly collapsing cloud of
self-gravitating dust particles has been examined.

An unresolved issue of cosmology is, the cosmic baryon asymmetry, predict a modification of general relativity
with the insertion of a Chern-Simons (CS) correction during the inflationary period \cite{[28]}.
The CS modified gravity is an extension to GR. The metric is coupled to a
scalar field which yields the modified Einstein field equations, also called the field equations of CS modified gravity.

A number of black hole solutions satisfying the modified equations of gravity have been discussed by
Grumiller and Yunes \cite{[29]}. The Schwarzschild solution is harmonious with CS modified gravity studied
in \cite{[30]}. Furtado with his collaborators investigated the consistency of the G\"{o}del metric, both
for external and dynamical CS coefficients given in \cite{[31]}. The solution of one parameter
family of G\"{o}del-type metrics has been proved consistent by Ahmedov and Aliev \cite{[32]}. The birefringence
of the gravitational waves is discussed in \cite{[30]}, some cosmological effects have been investigated in
\cite{[34],[35]}, the post-Newtonian expansion is considered in \cite{[36]}. Recently, Jamil and Sarfraz \cite{[47]} evaluated
the Ricci dark energy of amended FRW model in the framework of CS modified gravity and discuss it graphically.
In this paper we discuss the spherical symmetric gravitational collapse in the framework of dynamical and non-dynamical
CS modified gravity.

This paper is organized in following order. The junction conditions are investigated in
section $2$. In section $3$, we found the solution of the field equations in the framework of CS modified gravity.
The apparent horizons are studied in section $4$. The results are concluded in the last section.

\section{Junction Conditions}
This section provides the study of the junction conditions at the surface of a collapsing sphere with dust fluid.
We assume 4D spherically symmetric spacetime about an origin  $O$ and 3D hypersurface $\Sigma$ centered at
$O$ which divides spherically symmetric spacetime into two regions, namely, interior region $V^{-}$ and
exterior region $V^{+}$.

The interior region is represented by the line element
\begin{eqnarray}
ds^{2}_{-}=dt^{2}-X^{2}dr^{2}-Y^{2}(d\theta^{2}+\sin^{2}{\theta}d\phi^{2}),
\end{eqnarray}
where X and Y are functions of t and r. For exterior region, we consider the Schwarzschild spacetime which is described as
\begin{eqnarray}
ds^{2}_{+}=(1-\frac{2M}{R})dT^{2}-\frac{1}{(1-\frac{2M}{R})}dR^{2}-R^{2}(d\theta^{2}+\sin^{2}{\theta}d\phi^{2}),
\end{eqnarray}
where M is constant.

Using the following junction conditions:\\
1. The continuity of line element over $\Sigma$ yields
\begin{eqnarray}
(ds^{2}_{+})_{\Sigma}= (ds^{2}_{-})_{\Sigma} = (ds^{2})_{\Sigma}.
\end{eqnarray}
2. The continuity of extrinsic curvature over $\Sigma$ provides
\begin{eqnarray}
[K_{ij}]= K^{+}_{ij}-K^{-}_{ij}=0,~~~~(i,j= 0,2,3)
\end{eqnarray}
 where $K_{ij}$ is the extrinsic curvature tensor defined as
\begin{eqnarray}
K^{\pm}_{ij}=-n^{\pm}_{\sigma}(\frac{\partial^{2}x^{\sigma}_{\pm}}{\partial \varepsilon^{i}\partial \varepsilon^{j}}+
\Gamma^{\sigma}_{\mu\nu}\frac{\partial x ^{\mu}_{\pm}}{\partial \varepsilon^{i}} \frac{\partial x ^{\nu}_{\pm}} {\partial \varepsilon^{j}})=0.~~~~(\sigma,\mu,\nu= 0,1,2,3)
\end{eqnarray}
Here  $n^{\pm}_{\sigma}$ are components of outword unit normal to the hypersurface $\Sigma$ in coordinates
$x^{\sigma}_{\pm}$ of $V^{\pm}$, $\varepsilon^{i}$ correspond to
the coordinates on $\Sigma$ and $\Gamma^{\sigma}_{\mu\nu}$ are evaluated for the interior and exterior spacetimes.

Using interior and exterior spacetimes coordinates, the equations of hypersurface are calculated as
 \begin{eqnarray}
h_{-}(r,t)&=&r-r_{\Sigma}=0,\\
h_{+}(R,T)&=&R-R_{\Sigma}(T)=0,
\end{eqnarray}
where $r_{\Sigma}$ is an arbitrary constant.
Using above expressions in Eq.(1) and Eq.(2), the corresponding equations are given by.

\begin{eqnarray}
ds^{2}_{-}&=&dt^{2}-Y^{2}(r_{\Sigma},t)(d\theta^{2}+\sin^{2}{\theta}d\phi^{2}),\\
ds^{2}_{+}&=&[1-\frac{2M}{R_{\Sigma}}-\frac{1}{1-\frac{2M}{R_{\Sigma}}}(\frac{dR_{\Sigma}}{dT})^{2}]dT^{2}
-R^{2}_{\Sigma}(d\theta^{2}+\sin^{2}{\theta}d\phi^{2}).
\end{eqnarray}

We suppose that
\begin{eqnarray}
(1-\frac{2M}{R_{\Sigma}}-\frac{1}{1-\frac{2M}{R_{\Sigma}}}(\frac{dR _{\Sigma}}{dT})^{2})&>&0,
\end{eqnarray}
so $T$ remains timelike coordinate. Using junction condition given in Eq.(3), we obtain
\begin{eqnarray}
R_{\Sigma}=Y(r_{\Sigma},t),\\
(1-\frac{2M}{R_{\Sigma}}-\frac{1}{1-\frac{2M}{R_{\Sigma}}}(\frac{dR _{\Sigma}}{dT})^{2})^{\frac{1}{2}}dT=dt.
\end{eqnarray}
The outward unit normals in interior region $V^{-}$ and in exterior region $V^{+}$ are given by Eq.(6) and (7)
\begin{eqnarray}
n_{\mu}^{-}&=&(0,X(r_{\Sigma},t),0,0)\\
n_{\mu}^{+}&=&(-\dot{R_{\Sigma}},\dot{T},0,0).
\end{eqnarray}
Using Eq.(5), the components of the extrinsic curvature $K^{\pm}_{ij}$ has been calculated as
\begin{eqnarray}
K^{-}_{00}&=&0\\
K^{-}_{22}&=&csc^{2}\theta K^{-}_{33}=(\frac{YY^{\prime}}{X})_{\Sigma},\\
K^{+}_{00}&=&(\dot{R}\ddot{T}-\dot{T}\ddot{R}+\frac{3M\dot{R}^{2}\dot{T}}{R(R-2M)}-\frac{M(R-2M)\dot{T}^{3}}{R^{3}})_{\Sigma}, \\
K^{+}_{22}&=&csc^{2}\theta K^{+}_{33}=[\dot{T}(R-2M)]_{\Sigma},
\end{eqnarray}
here dot and prime stand for differentiation with respect to $t$
and $r$ respectively. According to continuity conditions of extrinsic
curvature,
\begin{eqnarray}
K^{+}_{00}&=&0, ~~ K^{+}_{22}=K^{-}_{22}.
\end{eqnarray}
Utilizing the Eqs.(11)-(12) and (15)-(18) in Eq.(19), we arrive at
\begin{eqnarray}
(X \dot{Y}^{\prime}  - \dot{X}Y^{\prime})_{\Sigma}=0, \\
M=(\frac{Y}{2}+\frac{Y}{2}\dot{Y}^{2}-\frac{Y}{2X^{2}}Y^{{\prime}2})_{\Sigma}.
\end{eqnarray}.

\section{Brief Review of CS Modified Gravity}
The Einstein-Hilbert action is modified by adding the gravitational CS term, is given as
\begin{eqnarray}
S=\int d^{4}x\sqrt{-g}[\kappa R+\frac{\alpha}{4}\Theta~^{*}RR-\frac{\beta}{2}(g^{\mu\nu}\nabla_{\mu}\Theta\nabla_{\nu}\Theta+2V[\Theta])]+S_{mat},
\end{eqnarray}
where $\kappa^{-1} = 16\pi G$, $\alpha$ and $\beta$ are coupling constants, $g$ is the determinant
of the metric, $\nabla_{\mu}$ is the covariant derivative associated with $g_{ab}$, $R$ is the Ricci scalar, $^{*}RR$
is called Pontryagin term which is topological invariant, $S_{mat}$ is matter action and the function $\Theta$ is the
so-called CS coupling field.  It is mentioned here that the function $\Theta$ may not be taken as a constant, but a function of spacetime, thus serving as a deformation function. Otherwise CS modified gravity reduces identically to GR. The Pontryagin term is defined as
\begin{eqnarray}
^{*}RR= {{^{*}R^a}_b}^{cd} {R^b}_{acd},
\end{eqnarray}
where ${R^b}_{acd}$ is the Reimann tensor and ${{^{*}R^a}_b}^{cd}$ is the dual Reimann tensor defined as
\begin{eqnarray}
{{^{*}R^a}_b}^{cd}=\frac{1}{2}\epsilon^{cdef}{R^a}_{bef}.
\end{eqnarray}
Now, the variation of the action given in Eq.(22) with respect to scalar field $\Theta$ and the metric tensor $g_{\mu\nu}$,
the two field equations of CS gravity, called modified Einstein field equations (EFEs) are given by
\begin{eqnarray}
G_{\mu\nu}+l C_{\mu\nu} &=& \kappa T_{\mu\nu},\\
\beta \square \Theta &=&\beta\frac{dV}{d\Theta}-\frac{\alpha}{4}~{^{*}RR}.
\end{eqnarray}

where $G_{\mu\nu}$ is the Einstein tensor, $C_{\mu\nu}$ is the C-tensor and  $T_{\mu\nu}$  is energy-momentum tensor.
In the context of classical and semi-classical scenarios of String theory, the potential $V[\Theta]$ is negligible.
Energy-momentum tensor is consist on two parts, one is matter part $T^{m}_{\mu\nu}$ and other is external
field part $T^{\Theta}_{\mu\nu}$ defined respectively as
\begin{eqnarray}
T^{m}_{\mu\nu}&=&(\rho+p)U_\mu U_\nu-p g_{\mu\nu},\\
T^{\Theta}_{\mu\nu}&=&\beta(\partial_{\mu}\Theta)(\partial_{\nu}\Theta)-\frac{\beta}{2}g_{\mu\nu}(\partial^\lambda\Theta)(\partial_{\lambda}\Theta),
\end{eqnarray}
where $\rho$ is energy density, $p$ is pressure and $U$ four-vector velocity in co-moving
coordinates of the spacetime. The C-tensor is expressed as
\begin{eqnarray}
C^{\mu\nu}=-\frac{1}{2\sqrt{-g}}[\upsilon_{\sigma}\epsilon^{\sigma\mu\alpha\beta}\nabla_{\alpha}R^{\nu}_{\beta}+
\frac{1}{2}\upsilon_{\sigma\tau}\epsilon^{\sigma\nu\alpha\beta}R^{\tau\mu}_{\alpha\beta}]+(\mu\longleftrightarrow\nu),
\end{eqnarray}
where
\begin{eqnarray}
\upsilon_{\sigma}\equiv\nabla_{\sigma}\Theta,~~~~~ \upsilon_{\sigma\tau}\equiv\nabla_{\sigma}\nabla_{\tau}\Theta.
\end{eqnarray}
The vanishing of the C-tensor depends on the type of relation between the Ricci tensor and Levi-Civita tensor.\\

The CS modified theory has been divided into two distinct formulations named as dynamical and non-dynamical.\\
\textbf{In non-dynamical case} $\beta = 0$ is considered in field equations given in Eq. (25) and (26), we get
\begin{eqnarray}
G_{\mu\nu}+l C_{\mu\nu} &=& \kappa T^{m} _{\mu\nu},\\
0&=&~{^{*}RR}.
\end{eqnarray}
If the vacuum case is under consideration, the right-hand side of the first equation become identically zero.
The second equation which is used for the evaluation equation of $\Theta$, is another differential constraint in
the allowed solution space. In non-dynamical case the canonical choice of $\Theta$ is chosen as, postulated in \cite{[30]}
\begin{eqnarray}
\Theta&=& \frac{t}{\mu}.
\end{eqnarray}
\textbf{In dynamical formulation} the $\beta$ is taken arbitrary and the modified field equations are given by
Eqs. (25)-(30).

The well known Einstein tensor $G_{\mu\nu}$ is written as
\begin{eqnarray}
G_{\mu\nu}&=&R_{\mu\nu}-\frac{1}{2}g_{\mu\nu}R.
\end{eqnarray}

Making use of Eqs.(27), (28) and (34) in Eq.(25), the trace-reversed form of
the CS modified field equations has been evaluated in dust case, setting $\beta=1$, similar to \cite{[40]}, as
\begin{eqnarray}
R_{\mu\nu}&=&\kappa[\rho U_{\mu}U_{\nu}-\frac{1}{2}\rho g_{\mu\nu}-(\partial_{\mu}\Theta)(\partial_{\nu}\Theta)]-l C_{\mu\nu}.
\end{eqnarray}
The evolution equation of $\Theta$ is given by
\begin{eqnarray}
g^{\mu \nu}\nabla_{\mu}\nabla_{\nu}\Theta &=&-\frac{\alpha}{4}~^{*}RR
\end{eqnarray}
\section{Solutions of Modified Field Equations}
In the framework of dynamical CS modified gravity, the modified EFEs corresponding to spherical symmetric spacetime given in Eq.(1) yield
\begin{eqnarray}
&&-\frac{\ddot{X}}{X}-2\frac{\ddot{Y}}{Y}=4\pi \rho-8\pi(\partial_{0}\Theta)^{2},\\
&&\ddot{X}{X}+\frac{2X\dot{X}\dot{Y}}{Y}-\frac{2Y^{\prime\prime}}{Y}-\frac{2X^{\prime}Y^{\prime}}{{XY}}=4\pi\rho X^{2}+8\pi(\partial_{1}\Theta)^{2},\\
&&\ddot{Y}Y+\frac{Y\dot{Y}\dot{X}}{X}+\dot{Y}^{2}-\frac{Y Y^{\prime\prime}}{X^{2}}+\frac{Y X^{\prime}Y^{\prime}}{X^{3}}-\frac{Y^{\prime^2}}{X^{2}}+1\nonumber\\
&=&4\pi\rho Y^{2}+8\pi(\partial_{2}\Theta)^{2},\\
&&\sin^{2}\theta(\ddot{Y}Y+\frac{Y\dot{Y}\dot{X}}{X}+\dot{Y}^{2}-\frac{Y Y^{\prime\prime}}{X^{2}}+\frac{Y X^{\prime}Y^{\prime}}{X^{3}}-\frac{Y^{\prime^2}}{X^{2}}+1)\nonumber\\
&=&4\pi \rho Y^{2}\sin^{2}\theta+8\pi(\partial_{3}\Theta)^{2},\\
&&\frac{2\dot{X}Y^{\prime}}{XY}-\frac{2\dot{Y^{\prime}}}{Y}=8\pi(\partial_{0}\Theta)(\partial_{1}\Theta),\\
&&8\pi(\partial_{0}\Theta)(\partial_{2}\Theta)+\frac{l}{2X^5 Y^4} \csc\theta(-3\partial_{3}\Theta Y X^{\prime 2}Y^{\prime} + X(-\partial_{3} \Theta X^{\prime}Y^{\prime 2}\nonumber\\
&+& Y(\partial_{3} \Theta Y^{\prime}X^{\prime\prime}+X^{\prime}( \partial_{3}\partial_{1}\Theta Y^{\prime}+ 3\partial_{3} \Theta Y^{\prime\prime})))+ X^2(\partial_{3}\partial_{1}\Theta Y^{\prime 2}\nonumber\\
 &+& \partial_{3} \Theta Y^{\prime}X^{\prime\prime}+Y(- \partial_{3}\partial_{1}\Theta Y^{\prime\prime} - \partial_{3} \Theta(Y^{\prime\prime\prime}+ X^{\prime}(\dot{X}\dot{Y} - Y\ddot{X}))))\nonumber\\
&+& X^{3}(\partial_{3} \Theta Y^{\prime}\dot{X}\dot{Y}+Y(\dot{X}(\partial_{3}\partial_{1}\Theta\dot{Y} + \partial_{3} \Theta\dot{Y^{\prime}}) + \partial_{3} \Theta(\dot{Y}\dot{X}^{\prime} - 2Y^{\prime}\ddot{X}))\nonumber\\
&-& Y^{2}(\partial_{3}\partial_{1}\Theta\ddot{X} + \partial_{3} \Theta\ddot {X^{\prime}}))+ X^{4}(- \partial_{3}\partial_{1}\Theta - \partial_{3}\partial_{1}\Theta\dot{Y}^{2} - 2\partial_{3} \Theta\dot{Y}\dot{Y}^{\prime}\nonumber\\
&+&\partial_{3} \Theta Y^{\prime}\ddot{Y}+ Y(\partial_{3}\partial_{1}\Theta\ddot{Y} + \partial_{3} \Theta\ddot{Y}^{\prime})))=0,\\
&&8\pi(\partial_{0}\Theta)(\partial_{3}\Theta)+\frac{l}{2X^5Y^4} \csc\theta(- 3\partial_{2} \Theta YX^{\prime 2}Y^{\prime} + X(-\partial_{2} \Theta X^{\prime}Y^{\prime 2}\nonumber\\
&+&Y(\partial_{2} \Theta Y^{\prime}X^{\prime\prime}+X^{\prime}( \partial_{2}\partial_{0}\Theta Y^{\prime}+3\partial_{2} \Theta Y^{\prime\prime})))+ X^2(\partial_{2}\partial_{0}\Theta Y^{\prime 2} + \partial_{2} \Theta Y^{\prime}X^{\prime\prime}\nonumber\\
&+&Y(- \partial_{2}\partial_{0}\Theta Y^{\prime\prime} - \partial_{2} \Theta(Y^{\prime\prime\prime}+ X^{\prime}(\dot{X}\dot{Y} - Y\ddot{X}))))+ X^{3}(\partial_{2} \Theta Y^{\prime}\dot{X}\dot{Y}\nonumber\\
&+&Y(\dot{X}(\partial_{2}\partial_{0}\Theta\dot{Y}+\partial_{2} \Theta\dot{Y^{\prime}}) + \partial_{2} \Theta(\dot{Y}\dot{X}^{\prime}-2Y^{\prime}\ddot{X}))- Y^{2}(\partial_{2}\partial_{0}\Theta\ddot{X} + \partial_{2} \Theta\ddot {X^{\prime}}))\nonumber\\
&+&X^{4}(- \partial_{2}\partial_{0}\Theta-\partial_{2}\partial_{0}\Theta\dot{Y}^{2} - 2\partial_{2} \Theta\dot{Y}\dot{Y}^{\prime}+ \partial_{2} \Theta Y^{\prime}\ddot{Y}\nonumber\\
&+& Y(\partial_{2}\partial_{0}\Theta\ddot{Y} + \partial_{2} \Theta\ddot{Y}^{\prime})))=0,\\
&&(\partial_{1}\Theta)(\partial_{2}\Theta)=0,\\
&&(\partial_{1}\Theta)(\partial_{3}\Theta)=0,\\
&&(\partial_{2}\Theta)(\partial_{3}\Theta)=0
\end{eqnarray}
To evaluate $\Theta$, we use Eq. (36). The Pontryagin term  for spherical symmetric spacetime turned to be zero. Since $\Theta$ is a function of spacetime coordinates. For convenience, we consider $\Theta$ depend upon radial coordinate $r$ only, using the evolution equation Eq.(36), it turned to be
\begin{eqnarray}
\partial_{1}\Theta&=& \frac{S^{2}X^{2}}{Y^{2}},
\end{eqnarray}
where $S$ is constant of integration.
Substituting this value in Eq.(37)-(46), these equations are reduced in new set of field equations.
\begin{eqnarray}
&&\frac{\ddot{X}}{X}+2\frac{\ddot{Y}}{Y}=4\pi\rho,\\
&&\frac{\ddot{X}}{X}+\frac{2\dot{X}\dot{Y}}{XY}-\frac{2}{X^{2}}(\frac{{Y^{\prime\prime}}}{Y}-\frac{X^{\prime}Y^{\prime}}{{XY}})=4\pi\rho-\frac{8\pi S^{2}}{Y^{4}},\\
&&\frac{\ddot{Y}}{Y}+(\frac{\dot{Y}}{Y})^{2}+\frac{\dot{X}\dot{Y}}{XY}-\frac{1}{X^{2}}[\frac{Y^{\prime\prime}}{Y}+(\frac{Y^{\prime}}{Y})^{2}
-\frac{X^{\prime}Y^{\prime}}{XY}-(\frac{X}{Y})^{2}]\nonumber\\
&&=4\pi\rho,\\
&&\frac{2\dot{X}Y^{\prime}}{XY}-\frac{2\dot{Y^{\prime}}}{Y}=0.
\end{eqnarray}

The integration of Eq.(51) yields
\begin{eqnarray}
X&=&\frac{Y^{\prime}}{W},
\end{eqnarray}
where $W=W(r)$ is arbitrary function of $r$. The result obtained in Eq.(52) is same as in Eq.(20) of junction condition, coincidentally.
Using in Eq.(48)-(50),
it follows that
\begin{eqnarray}
\frac{2\ddot{Y}}{Y}+(\frac{\dot{Y}}{Y})^{2}+\frac{1-W^{2}}{Y^{2}}&=&\frac{4\pi S^{2}}{Y^{4}}.
\end{eqnarray}
Integrating the above equation w.r.t $t$, we evaluated that
\begin{eqnarray}
\dot{Y}^{2}&=&W^{2}-1+\frac{2m}{Y}-\frac{4\pi S^{2}}{Y^{2}},
\end{eqnarray}
where $m$ is arbitrary function of $r$ and related to the mass of the
collapsing system. Using Eq.(54) and Eq.(52) in Eq.(48), we have
\begin{eqnarray}
m^{\prime}&=&8\pi\rho Y^{2}Y^{\prime}+20\pi S^{2}\frac{Y^{\prime}}{Y^{2}}.
\end{eqnarray}
Integrating w.r.t $r$ we get,
\begin{eqnarray}
m&=&4\pi\int^{r}_{0}(2\rho WXY^{2}+\frac{5S^{2}WX}{Y^{2}})dr +q(t).
\end{eqnarray}
where $q(t)$ is function of integration.
Now, using Eqs.(52) and (54) in junction condition (21), it turns out
\begin{eqnarray}
M&=&m-\frac{2\pi S^{2}}{Y^{2}}.
\end{eqnarray}
The definition of mass function given by Misner and Sharp \cite{[41]} can be used to calculate the
total energy $\tilde{M}(r,t)$ of the system up to radius $r$ at time $t$ inside the hypersurface $\Sigma$
\begin{eqnarray}
\tilde{M}(r,t)&=&\frac{1}{2}Y(1+g^{\mu\nu}Y_{,\mu}Y_{,\nu}).
\end{eqnarray}
For the interior metric, the total energy become
\begin{eqnarray}
\tilde{M}(r,t)&=&\frac{1}{2}Y(1+\dot{Y}-\frac{Y^{\prime ^2}}{X^{2}}).
\end{eqnarray}
Substituting the corresponding values, we get
\begin{eqnarray}
\tilde{M}(r,t)&=&m-\frac{2\pi S^{2}}{Y^{2}}.
\end{eqnarray}
The analytic solution, in closed form has been obtained, making the use of Eq.(52) in Eq.(54) along with the assumption $4 \pi S^{2}>0$ and $W=1$, as
\begin{tiny}
\begin{eqnarray}
Y&=&-\frac{2S^{2}\pi}{m} \nonumber\\
&-&\frac{4\times 2^{2/3} S^4 \pi ^2 m}{[-32 S^6 \pi ^3 m^6+
3 \sqrt{64 S^6 \pi ^3 m^{16} (t-t_s)^2+9m^{20}(t-t_s)^{4}}-9 m^{10}(t-t_s)^2]^{1/3}}\nonumber\\
&-&\frac{[-32 S^6 \pi ^3 m^6 + 3\sqrt{64 S^6 \pi ^3 m^{16} (t-t_s)^2+ 9m^{20}(t-t_s)^{4}}-9 m^{10}(t-t_s)^2]^{1/3}}{2^{2/3}m^{3}}.\nonumber\\
\end{eqnarray}
\end{tiny}
\begin{tiny}
\begin{eqnarray}
X&=& \frac{2S^{2}m^{\prime}}{m^{2}}\nonumber\\
&-&\frac{4\times 2^{2/3} S^4 \pi ^2 m^{\prime}}{[-32 S^6 \pi ^3 m^6+
3 \sqrt{64 S^6 \pi ^3 m^{16} (t-t_s)^2+9m^{20}(t-t_s)^{4}}-9 m^{10}(t-t_s)^2]^{1/3}}\nonumber\\
&-&\frac{3m^{\prime}[-32 S^6 \pi ^3 m^6 + 3\sqrt{64 S^6 \pi ^3 m^{16} (t-t_s)^2+ 9m^{20}(t-t_s)^{4}}-9 m^{10}(t-t_s)^2]^{1/3}}{2^{2/3}m^{4}}\nonumber\\
&+&8\times2^{2/3} S^4\pi^{2} m^6[-32 S^6 \pi ^3 m^{\prime}-15 m^4 (t-t_s)^2 m^{\prime}+3 m^5 (t-t_s) t_s^{\prime}\nonumber\\
&-&\frac{(m^{10} (t-t_s)(-(256S^6 \pi ^3+45 m^4 (t-t_s)^2)(t-t_s) m^{\prime}+m(32 S^6 \pi ^3+9 m^4(t-t_s)^2) t_s^{\prime}))}
{(\sqrt{64 S^6 \pi ^3 m^{16}(t-t_s)^2+9 m^{20}(t-t_s)^4})}]\nonumber\\
&\times&[-32 S^6 \pi^3 m^6-9 m^{10} (t-t_s)^2+3 \sqrt{m^{16} (64 S^6 \pi ^3+9 m^4(t-t_s)^2)(t-t_s)^2}]^{-4/3}\nonumber\\
&+&2^{1/3} m^2[-32 S^6 \pi ^3 m^{\prime}-15 m^4 (t-t_s)^2 m^{\prime}+3 m^5 (t-t_s) t_s^{\prime}\nonumber\\
&-&\frac{(m^{10} (t-t_s)(-(256S^6 \pi ^3+45 m^4 (t-t_s)^2)(t-t_s) m^{\prime}+m(32 S^6 \pi ^3+9 m^4(t-t_s)^2) t_s^{\prime}))}
{(\sqrt{64 S^6 \pi ^3 m^{16}(t-t_s)^2+9 m^{20}(t-t_s)^4})}]\nonumber\\
&\times&[-32 S^6 \pi^3 m^6-9 m^{10} (t-t_s)^2+3 \sqrt{m^{16} (64 S^6 \pi ^3+9 m^4(t-t_s)^2)(t-t_s)^2}]^{-2/3}
\end{eqnarray}
\end{tiny}
Here $t_{s}$ is an arbitrary function of $r$ and it represents the time formation of singularity for the
particular shell at radial distance $r$.
On limiting approach $ 4\pi S^{2}\rightarrow0$, the above solution takes the form
\begin{eqnarray}
\lim_{ 4\pi S^{2}\rightarrow0}Y&=&[\frac{9m}{2}(t_{s}(r)-t)^{2}]^{\frac{1}{3}},
\\
\lim_{ 4\pi S^{2}\rightarrow0}X&=& \frac{m^{\prime}(t_{s}(r)-t)+2mt^{\prime}_{s}(r)}{[6m^{2}(t_{s}(r)-t)]^{\frac{1}{3}}}.
\end{eqnarray}
these solutions correspond to the well known Tolman-Bondi solution given in \cite{[37]}\\

\textbf{In non-dynamical case} we discuss the solution of the spherical symmetric spacetime. As we examine that the Pontryagin term $^{*}RR=0$, so the evaluation equation is disappeared. Now making the canonical choice for the value of $\Theta=\frac{t}{\mu}$, the term $C_{\mu \nu}$ turned to be zero. So, the modified field equations reduces to general field equations of general reltivity.
\section{Apparent Horizon}

The nature of the singularities can be determined using the apparent horizons which are the
boundaries of the trapped regions. If the apparent horizon occurs before the formation of the
singularity then it is black hole and if the order is reversed then the singularity is said to be naked.
In four-dimensional spacetime, the formation of boundary of trapped two spheres becomes the cause of occurrence of the apparent horizon.
Now, we explore the boundary of trapped two spheres whose outword normals are null. Using null condition,
for the interior region, it yields
\begin{eqnarray}
g^{\mu\nu}Y_{,\mu}Y_{,\nu}&=&\dot{Y}^{2}-(\frac{Y^{\prime}}{X})^{2}=0.
\end{eqnarray}
Substituting the values of $\dot{Y}$ and $Y^{\prime}$ in Eq.(65), we arrive at
\begin{eqnarray}
Y^{2}-2mY+4 \pi S^{2}=0,
\end{eqnarray}
which is a quadratic equation in Y. Its positive roots will yield the apparent horizons. Now,
we discuss the different cases of the Eq.(66). Straightforwardly, Eq.(66) yields the
Schwarzschild horizons, i.e., $Y=2m$,  by using the assumption of $4 \pi S^{2}=0$.
For the case when $4 \pi S^{2}>0$, there arise three possibilities according as:
\begin{eqnarray}
1.~~~~~~~~m&>&\sqrt{4 \pi S^{2}},\nonumber\\
2.~~~~~~~~m&=&\sqrt{4 \pi S^{2}},\nonumber\\
3.~~~~~~~~m&<&\sqrt{4 \pi S^{2}}.\nonumber\\
\end{eqnarray}
\textbf{Case 1}. When $m>\sqrt{4 \pi S^{2}}$, we obtain two horizons,
one is cosmological horizon and other is called black hole horizon, denoted by $Y_{c}$ and $Y_{bh}$ respectively. These are given by
\begin{eqnarray}
Y_{c}=m+\sqrt{m^{2}-4 \pi S^{2}},
\end{eqnarray}
\begin{eqnarray}
Y_{bh}=m-\sqrt{m^{2}-4 \pi S^{2}},
\end{eqnarray}
If $m=0$ then there is no horizons.
If $m\neq0$ and $m\neq\sqrt{4 \pi S^{2}}$ then $Y_{c}$ and $Y_{bh}$ can be easily generalized by following \cite{[42]}.

\textbf{Case 2.}
For $m=\sqrt{4 \pi S^{2}}$, there is only one positive root, so we have a single horizon, i.e.,
\begin{eqnarray}
Y_{c}=Y_{bh}=m.
\end{eqnarray}
It means that both the horizons coincide with each other. It is worth mentioning here that it is again like a Schwarzschild horizon. The range for the cosmological horizon and black hole horizon is calculated as
\begin{eqnarray}
0\leq Y_{bh}\leq (m-\sqrt{m^{2}-4 \pi S^{2}})\leq Y_{c}\leq (m+\sqrt{m^{2}-4 \pi S^{2}}).
\end{eqnarray}
 The largest proper area of the black hole horizon is $4\pi Y^{2}$, calculated as
\begin{eqnarray}
4 \pi(m-\sqrt{m^{2}-4 \pi S^{2}})^{2}
\end{eqnarray}
and the cosmological horizon has its area between
\begin{eqnarray}
4 \pi(m-\sqrt{m^{2}-4 \pi S^{2}})^{2}~~~~~~\texttt{and}~~~~~~4 \pi(m+\sqrt{m^{2}-4 \pi S^{2}})^{2}.
\end{eqnarray}

\textbf{Case 3.} For $m<\sqrt{4 \pi S^{2}}$, there is no positive root. So, there is no apparent horizon.

Now, we discuss the formation time for the apparent horizon and singularities. There are two possibilities arise in the formation time for apparent horizon and singularity which can be calculated using Eq.(61) in Eq.(66).\\
\textbf{P-1}
\begin{eqnarray}
t_{n}=t_{s}-\frac{\sqrt{2}}{3}\sqrt{-\frac{24 S^4 \pi ^2}{m^2}-\frac{32 S^6 \pi ^3}{m^4}+\frac{12 S^2 \pi  Y_{n}}{m}\pm\frac{\sqrt{-(4
S^2 \pi -2 m Y_{n})^3}}{m}},\nonumber\\ n=1,2.
\end{eqnarray}
\textbf{P-2}
\begin{eqnarray}
t_{n}=t_{s}+\frac{\sqrt{2}}{3}\sqrt{-\frac{24 S^4 \pi ^2}{m^2}-\frac{32 S^6 \pi ^3}{m^4}+\frac{12 S^2 \pi  Y_{n}}{m}\pm\frac{\sqrt{-(4
S^2 \pi -2 m Y_{n})^3}}{m}},\nonumber\\ n=1,2.
\end{eqnarray}
In Eq.(74), if $4\pi S^{2}\rightarrow0$, the time for apparent horizon is given by
\begin{eqnarray}
t_{ah}=t_{s}-\frac{4}{3}m.
\end{eqnarray}
This result corresponds to the well known Tolman-Bondi solution \cite{[37]}.
We conclude that apparent horizon precedes the singularity by an amount of co-moving time $\frac{4}{3}m$. This
shows that the singularity is covered, i.e., it is a black hole. It is mentioned here that the time difference
between the formation of apparent horizon and singularity is same as obtained in Tolmen-Bondi solution \cite{[37]}.
Further, from Eqs.(71) and (76), we conclude that $Y_{c}\geq Y_{bh}$ and $t_{2}\geq t_{1}$ respectively.
Here $t_{1}$ stand for the time formation of cosmological horizon and $t_{2}$ denotes the time formation
of black hole horizon. The inequality $t_{2}\geq t_{1}$ indicates that the cosmological horizon occurs
earlier than the black hole horizon.

In case of Eq.(75), the singularity occurs before the apparent horizon and yields a naked singularity consequently.

\section{Discussion}

The aim of this paper is to explore the different aspects of gravitational collapse in CS modified gravity
for spherical symmetric background. The modified EFE's are used to find the solutions for the particular
choice of the scalar field $\Theta$. The obtained solutions are compatible with the existing solutions, available
in literature, in the framework of GR \cite{[9]}-\cite{[16]}  and f(R) \cite{[25],[26]} gravities.

The relation for the Newtonian potential is given by $\Phi=\frac{1}{2}(1-g_{00})$. The use of Eq.(11)
in Eq.(57) for the exterior spacetime gives the Newtonian potential as
\begin{eqnarray}
\Phi(R)=\frac{m}{R}-\frac{2\pi S^{2}}{R^{2}}.
\end{eqnarray}
The Newtonian force corresponding to the Newtonian potential is obtained by taking derivative of the above equation w.r.t. $R$
\begin{eqnarray}
F=-\frac{m}{R^{2}}+\frac{4\pi S^{2}}{R^{3}}.
\end{eqnarray}
Now, we discuss the effect of Newtonian force on gravitational collapse.
This force has no effect on the collapsing process for the particular choice of the values of $m=\sqrt[3]{16\pi^{2} S^{4}}$ and
$R= \sqrt[3]{4\pi S^{2}}$. If the values of $m$ and $R$ are greater than the above defined
values then it is repulsive force, provided that $4\pi S^{2}$ is non zero.

The rate of gravitational collapse can calculated from Eq.(54) as
\begin{eqnarray}
\ddot{Y}=-\frac{2m}{Y^{2}}+\frac{8\pi S^{2}}{Y^{3}}.
\end{eqnarray}
For the gravitational collapse, the force must be attractive which indicates that the acceleration should be negative, i.e.,
$Y<\frac{4\pi S^{2}}{m}$.
Hence, if $ 4\pi S^{2}>0$ then the process of collapse will slow down.

Furthermore, we have calculated two apparent horizons named as cosmological horizon and black hole horizon.
It has been shown that the black hole horizon requires more time than cosmological horizon for its formation.
However, in \textbf{P-1}, both the horizons occur earlier than singularity forms. It means singularity is covered and
CCH \cite{[5]} is also supported by CS modified gravity.
In case \textbf{P-2}, the singularity occurs before the apparent horizons, that is, results a naked singularity.

\vspace{0.5cm}

{\bf Acknowledgment}
We acknowledge the remarkable assistance of the Higher Education
Commission Islamabad, Pakistan, and thankful for its financial
support through the {\it Indigenous PhD 5000 Fellowship Program
Batch-III}.

\end{document}